\documentclass[12pt]{iopart}
\begin{document}

\title[Photon deflection and  precession of the periastron]{Photon deflection and  precession of the periastron in terms of spatial gravitational fields}

\author{W.D.Flanders\dag\  and G.S.Japaridze\ddag\   \footnote[3]{To
whom correspondence should be addressed (japar@ctsps.cau.edu)}
}

\address{\dag\ Departments of Epidemiology and Biostatistics, Emory University, Atlanta, GA 30322, USA}

\address{\ddag\ Center for Theoretical Studies of Physical Systems, Clark Atlanta University, Atlanta, GA 30314, USA}

\begin{abstract}
We show that a Maxwell-like system of equations for  spatial gravitational fields $\bf  g$ and $\bf B$ (latter being the analogy of a magnetic field), modified to include an extra term for the $\bf B$ field in the expression for force,
leads to the correct values for the photon deflection angle and for the precession of the periastron. 

\end{abstract}


\maketitle

\section{Introduction}
The attempts to describe gravitational phenomena in terms of spatial gravitational fields are known from the times of Maxwell \cite{holz}. After establishing the General Theory of Relativity it became clear that the vector theory could not serve as an adequate description of gravitational interactions since the linear equations can be viewed at best as some approximation of an adequate theory, presented by the tensor equations of General Relativity. 

In this paper we suggest a modified version of a Maxwell-like system of equations for the gravitational field,  calculate the deflection of a photon passing the central body and the precession of the planetary motion, and compare the results with those of General Relativity \cite{landau}, \cite{dirac}.

We postulate fields ${\bf g}$ and ${\bf B}$ satisfying the system of equations
\begin{eqnarray}
div\;{\bf g}=4\pi G\rho,\; curl\;{\bf g}=-{\partial {\bf B}\over {\partial {t}}}
\cr\cr
div\;{\bf B}\,=0,\; curl\;{\bf B}\,=\frac{1}{c^{2}}\frac{\partial {\bf g}}{\partial {t}}-\frac{4\pi\;G}{c^{2}}\bf j
\label{equations}
\end{eqnarray}
In (\ref{equations}), $G$ is Newton's constant, $c$ is the speed of light,  $\rho$ is the mass density and $\bf j$ is the mass density flow.

We also postulate that the gravitational force acting on a moving particle with velocity $\bf v$  and mass $m$ ($m\equiv m_{0}\,\gamma$, $\gamma^{-1}\equiv \sqrt{1-v^{2}/c^{2}}$)  is described by:
\begin{eqnarray}
\label{basicEq}
{\bf F}\,\equiv\,\frac{d{\bf p}}{dt}\,\equiv\,\frac{d(m {\bf v)}}{dt}\,\equiv\,m\,{\bf a}+
m\gamma^{2}\frac{({\bf v}{\bf a}){\bf v}}{c^{2}}\,=
\cr\cr
=\,m\,{\bf g}\, +\,m\,{\bf v}\,\times\,({\bf B}\,+\,\delta{\bf B})
\end{eqnarray}
where ${\bf a}\equiv d{\bf v}/dt$, and $\delta{\bf B}$ is defined by
\begin{equation}
\label{!!}
 curl\;({\bf B}\,+\,\delta{\bf B})\,=\frac{1}{c^{2}}\frac{d {\bf g}}{d {t}}-\frac{4\pi\;G}{c^{2}}\bf j
\end{equation}

Equations similar to (\ref{equations}) and  to (\ref{basicEq})  (without the $\delta{\bf B}$ term) appear in the numerous attempts to model gravitational force after the electromagnetic Lorentz force (see e.g. \cite{cataneo} - \cite{fj}).
The system of linear equations (\ref{equations}) for the effective spatial gravitational fields are not accounting for the self-interaction, which appears in the framework of General Relativity in a natural way, and which is essential for deriving the correct values for the observables such as the photon deflection and the perihelion precession \cite{landau} - \cite{dirac}. 

Introduction of the  $\delta{\bf B}$-term  via (\ref{!!}) effectively leads to an extra contribution to the ${\bf B}$ field   which, in turn, results in an additional contribution $m{\bf v}\times \delta{\bf B}$ to the force (everywhere below for notational simplicity we write  ${\bf B}$ in place of ${\bf B}\,+\, \delta{\bf B}$). 

 Based on equations (\ref{equations}) - (\ref{!!}), we calculate to the lowest order in $v^{2}/c^{2}$ the deflection of a photon passing a central body and the precession of the planetary motion.  It turns out that equations (\ref{equations}) - (\ref{!!}) yield results which coincide with those following  from the General relativity \cite{landau}, \cite{dirac},  
 suggesting that at least some observables can be calculated using fields ${\bf g}$ and ${\bf B}$.

In section {\bf 2} we derive expressions for the angle of the deflection of a photon as it passes a central body (throughout,  we assume that the central body is the sun and use a coordinate system in which the sun is at rest), and for the precession of the perihelion of a planet orbiting the sun.  In section {\bf 3} we discuss our results as well as the limitations and approximations of the approach. In the Appendix we demonstrate how the equations of planetary motion which follow from our formulation and which lead to the correct expression for the precession, also follow from General Relativity provided that higher orders of $v^{2}/c^{2}$ are neglected and characteristic distances are bigger than Schwarzschild radius.

\section{Deflection of the photon and the precession of the perihelion of a planet}
\subsection{Photon Deflection}
\label{int}
We consider a photon of energy $E$ moving  along the $y$ axis and passing the sun at a perpendicular distance $d$.  At time instant $t$ the angle from the $x$ axis to the photon is $\theta(t)$:
\begin{equation}
\label{angle}
\theta(t)\,=\,\arctan\,{c\,t\over d},\quad {d\theta(t)\over dt}\,=\,{c\over d}\,\cos^{2}\theta(t)
\end{equation}
The $x$-component of the ${\bf g}$ field is
\begin{equation}
\label{gx}
{\bf g}_{x}\,=\,-\,{G\,M\over R^{2}}\,\cos \theta\,=\,-\,{G\,M\over d^{2}}\,\cos^{3}\theta
\end{equation}
where $M$ is the mass of the sun and $R\,=\,d/\cos\theta$ is the distance from the photon to the sun.

From the equations (\ref{equations}) we obtain for the derivative of ${\bf g}$ \begin{equation}
\label{gderiv}
{1\over c^{2}}\,{d{\bf g}_{x}\over dt}\,=\,3\,{G\,M\over c\,d^{3 }}\,\cos^{4}\theta\,\sin\theta\,=\,(curl\,{\bf B})_{x}\,=\,{\partial {\bf B}_{z}\over \partial y}\,-\,{\partial {\bf B}_{y}\over \partial z}
\end{equation}
The value of $d\,{\bf g}/dt$ is not zero because we evaluate the change in ${\bf g}$ along the particle's path; in this case the non-trivial time dependence is realized by $\theta\,=\,\theta(t)$.

From the symmetry considerations it  follows that the $y$ component of the gravimagnetic field is vanishing: ${\bf B}_{y}\,=\,0$, so  integrating in (\ref{gderiv}) and using $\cos \theta\,=\,d/(d^{2}\,+\,y^{2})^{1/2}$ we obtain for ${\bf B}_{z}$:
\begin{equation}
\label{bz}
{\bf B}_{z}\,=\,\int^{c\,t}_{-\infty}\,{\partial {\bf B}_{z}\over \partial y}\,dy\,=\,-\,{G\,M\over c\,d^{2}}\,\cos^{3}\theta
\end{equation}
Now, having ${\bf g}_{x}$ and ${\bf B}_{z}$ at hand, we can evaluate the expression for ${\bf F}_{x}$
\begin{equation}
\fl
\label{forcex}
{\bf F}_{x}\,=\,m\,{\bf g}_{x}\,+\,m\,v_{y}\,{\bf B}_{z}\,=\,-\,{E\over c^{2}}\,{G\,M\over d^{2}}\,\cos^{3}\theta\,-\,{E\over c^{2}}\,c\,{G\,M\over c\,d^{2}}\,\cos^{3}\theta\,=\,-\,2\,{E\over c^{2}}\,{G\,M\over d^{2}}\,\cos^{3}\theta
\end{equation}
where we have used $m\,=\,E/c^{2}$ for the mass of the photon, and $v_{y}\,=\,c$ for its velocity.

The coefficient $2$ in (\ref{forcex}) is due to the contribution from the ``${\bf B}$'' - term; as is evident from (\ref{forcex}) the contributions from the ``${\bf g}$'' and the
 ``${\bf B}$'' terms are equal. In Newtonian physics ${\bf F}_{x}$  would have been $1/2$ of the value we obtained in (\ref{forcex}), and since the deflection angle is linear in ${\bf F}_{x}$ ( see below (\ref{px})), this  would result in the half of the deflection angle.

We now calculate the change of the $x$ component of the photon's momentum along its trajectory, $\delta {\bf p}_{x}$:
\begin{equation}
\label{px}
\delta {\bf p}_{x}\,=\,\int^{\infty}_{-\infty}\,dt\,{\bf F}_{x}\,=\,-\,{4\,G\,M\,E\over c^{3}\,d}
\end{equation}

Therefore,  for the value of the deflection angle $\delta\,\varphi$ defined from the expression $\tan(\delta\,\varphi)\,=\,|\delta {\bf p}_{x}/{\bf p}|$ we obtain
$\tan\,\delta\,\varphi\,=\,4\,G\,M/c^{2}\,d$, i.e. for small $\delta\,\varphi$ 
\begin{equation}
\label{deflection}
\delta\,\varphi\,=\,\arctan\,\Biggl( {4\,G\,M\over c^{2}\,d}\Biggr)\,\simeq \,{4\,G\,M\over c^{2}\,d}
\end{equation}
The same result  follows from the General Relativity \cite{landau}, \cite{dirac}.

\subsection{Precession of the Perihelion of Planetary Motion}
Let us consider the motion of a planet with  mass $m$ around the sun. At the moment $t$ we have for the planet ${\bf r}(t)\,=\,(x,\,y,\,z)\,=\,(R(\theta(t))\,\cos \theta(t),\,R(\theta(t))\,\sin \theta(t),\,0)$. 

For simplicity, we will present the calculations for the $x$ - component of the force and the acceleration at the point $\theta\,=\,0$. The results do not depend on this choice.

From (\ref{equations}) and ${\bf B}_{x}\,=\,0$ we obtain
\begin{equation}
\label{bzderivative}
{\partial {\bf B}_{z}\over \partial x}\,=\,{1\over c^{2}}\,{d {\bf g}_{y}\over d t}\,=\,{G\,M\over c^{2}}\,u^{2}\,\dot{\theta}
\end{equation}
where $M$ is the mass of the central body, $u\equiv 1/R$, and $\dot{\theta}\,=d\theta/dt$. Using $x\,=\,\cos\theta/u$ and integrating (\ref{bzderivative}) over $u$  leads to:
\begin{equation}
\label{anotherbz}
{\bf B}_{z}\,=\,-\,{G\,M\over c^{2}}\,u\,\dot{\theta}
\end{equation}
Combining (\ref{basicEq}), (\ref{anotherbz}), and $v_{y}\,=\,\dot{\theta}\,\cos\theta/u\,-\,u^{\prime}\,\dot{\theta}\,\sin\theta/u^{2}\,=\,\dot{\theta}/u$ (for $\theta\,=\,0$; $u^{\prime}\equiv du/d\theta$) we obtain for the $x$ - component:
\begin{equation}
\label{xdzala}
{\bf F}_{x}\,=\,-\,G\,M\,m\,u^{2}\,-\,{G\,M\over c^{2}}\,m\,\dot{\theta}^{2}
\end{equation}
Expanding in $v^{2}/c^{2}$ and retaining the first non-trivial terms, from ${\bf F}_{x}\,=\,d{\bf p}_{x}/dt$ and (\ref{xdzala}) it follows that
\begin{equation}
\label{xdzala1}
-\,G\,M\,m\,u^{2}\,-\,{G\,M\over c^{2}}\,m\,\dot{\theta}^{2}\,=\,\Biggl( 1\,+\,{v^{2}_{x}\over c^{2}}\Biggr)\,m\,a_{x}
\end{equation}
where $a_{x}$ is the $x$ component of the  planet's acceleration. In Newtonian physics and Special relativity, the term $m\,\dot{\theta}^{2}\,G\,M/ c^{2}$ would not appear.

Substituting $v_{x}\,=\,-\,u^{\prime}\,\dot{\theta}/u^{2}$ and  $a_{x}\,=\,-\,u^{\prime \prime}\,\dot{\theta}^{2}/u^{2}\,-\,u^{\prime}\,\ddot{\theta}/u^{2}\,+\,2\,(u^{\prime}\,\dot{\theta})^{2}/u^{3}\,-\,\dot{\theta}^{2}/u$ after  straightforward algebraic manipulations we get
\begin{equation}
\fl
\label{1}
-\,G\,M\,-\,{G\,M\,\dot{\theta}^{2}\over c^{2}\,u^{2}}\,+\,{G\,M\,(u^{\prime}\,\dot{\theta})^{2}\over c^{2}\,u^{4}}\,=\,-\,{u^{\prime \prime}\,\dot{\theta}^{2}\over u^{4}}\,-\,{u^{\prime}\,\ddot{\theta}\over u^{4}}\,+\,2\,{(u^{\prime}\,\dot{\theta})^{2}\over u^{5}}\,-\,{\dot{\theta}^{2}\over u^{3}}
\end{equation}
We can write expression (\ref{1}) in a more compact form by using $\bar{L}\equiv L/m\,=\,\dot{\theta}/u^{2}$  where $L$ is the angular momentum. We have
\begin{equation}
\label{2}
\fl
-\,G\,M\,-\,{G\,M\,u^{2}\,\bar{L}^{2}\over c^{2}}\,+\,{G\,M\,(u^{\prime}\,\bar{L})^{2}\over c^{2}}\,=\,-\, u^{\prime \prime}\,\bar{L}^{2}\,-\,{1\over 2}\,u^{\prime}\,{d\bar{L}^{2}\over d\theta}\,-\,u\,\bar{L}^{2}
\end{equation}

We will seek the solution of the equation of motion (\ref{2}) of the form $u\,=\,u_{0}\,+\,A\,\cos (k\,\theta\,+\,B)$. 

From equations (\ref{basicEq}) it follows that the $y$ - component of the force (for $\theta\,=\,0$) is non-vanishing, and is given by
\begin{equation}
\label{3}
{\bf F}_{y}\,=\,m\,{\bf B}_{z}\,v_{x}\,=\,{G\,M\,m\,u^{\prime}\over c^{2}}\,{\dot{\theta}^{2}\over u}
\end{equation}
We equate ${\bf F}_{y}$ to $d{\bf p}_{y}/dt$ and neglect the higher order terms in $1/c^{2}$ to obtain for $a_{y}\,=\,dv_{y}/dt$:
\begin{equation}
\label{4}
a_{y}\,=\,2\,{G\,M\, u^{\prime}\over c^{2}}\,{\dot{\theta}^{2}\over u}
\end{equation}
Note, that in the absence of the ${\bf B}$ field we would have ${\bf F}_{y}\,=\,0$, and as a consequence, $a_{y}$ would have been half as large.

Combining (\ref{4}), $a_{y}\,=\,u\,d\bar{L}/dt$, and ansatz $u\,=\,u_{0}\,+\,A\,\cos (k\,\theta\,+\,B)$ we obtain for $d\bar{L}^{2}/d\theta$
\begin{equation}
\label{5}
{d\,\bar{L}^{2}\over d\,\theta}\,=\,4{G\,M\, u^{\prime}\over c^{2}}\,{\dot{\theta}^{2}\over u^{4}}\,=\,4\,{G\,M \bar{L}^{2}\,A\over c^{2}}\,k\,\sin(k\,\theta\,+\,B)
\end{equation}
In approximation $GMA/c^{2}\,\ll \,1$ the solution of (\ref{5}) is
\begin{equation}
\label{6}
\bar{L}^{2}\,=\,\bar{L_{0}}^{2}\,\Biggl ( 1\,-\,4\,{GMA\over c^{2}}\,\cos(k\,\theta\,+\,B) \Biggr )
\end{equation}
Now we substitute $u\,=\,u_{0}\,+\,A\,\cos (k\,\theta\,+\,B)$ and (\ref{6}) into equation (\ref{2}), and  equate the same powers of $\cos (k\,\theta\,+\,B)$. The solutions for $u_{0}$ and $k^{2}$ are as follows:
\begin{equation}
\label{u}
u_{0}\,\simeq\,{G\,M\over \bar{L_{0}}^{2}},
\end{equation}
and
\begin{equation}
\label{k2}
k^{2}\,\simeq\,1\,-\,6\,{G^{2}\,M^{2}\over c^{2}\,\bar{L_{0}}^{2}}
\end{equation}
The corrections to (\ref{u}) - (\ref{k2})  are of higher order in $1/c^{2}$, and are numerically negligible.

Expression (\ref{k2}) for $k^{2}$ coincides with that from General Relativity \cite{landau}, \cite{dirac}, and of course, yields the correct result for the precession of the planetary perihelion. In particular, expression (\ref{k2}) gives the correct value for the observable precession of the perihelion of Mercury.

\section{Discussion}
We have demonstrated that the postulated approach,  equations (\ref{equations}) - (\ref{!!}),
leads to the correct expressions for the two key classic tests of General Relativity - the precession of  planetary motion and  the deflection of a photon.

The ${\bf B}$ field leads to a doubling of the force on the photon, and therefore to a doubling of the deflection obtained in the framework of Newtonian gravity.  This doubling leads to the correct result for the deflection.  

In the case of  planetary motion, the ${\bf B}$ field leads to an additional term in ${\bf F}_{x}$ and to a doubling of $a_{y}$ (see Equations (\ref{xdzala}) and (\ref{4})).  These values of ${\bf F}_{x}$ and $a_{y}$, used in the equation of motion (\ref {2}), lead to the correct expression for the perihelion precession.  If the ${\bf B}$ field were ignored in calculating  ${\bf F}_{x}$ and $a_{y}$, we would have obtained for $k^{2}$ the expression  $1\,-\,2\,G^{2}\,M^{2}/c^{2}\,\bar{L_{0}}^{2}$ instead of the correct value $k^{2}\,=\,1\,-\,6\,G^{2}\,M^{2}/c^{2}\,\bar{L_{0}}^{2}$. The observation that equations  (\ref{equations}) - (\ref{!!}) yield the correct results for these two different phenomena seems more than mere coincidence.

Let us address  an important question about our approach: what is the justification for the key postulates (\ref{equations}) - (\ref{!!}). In addition to the observation that they lead to the correct expressions for observables, we note that equations (\ref{xdzala}) and (\ref{3}) which are immediate results of our approach (\ref{equations}) - (\ref{!!}) follow from General Relativity, assuming spherical symmetry and time independence of the metric.
This derivation demonstrates consistency of our approach with General Relativity in case of weak fields and in order  $v^{2}/c^{2}$.
We sketch this derivation in the Appendix.

As an attempt to interpret our approach, we note that switching from a partial derivative to  a total derivative in ({\ref{!!}) which introduces an extra ${\bf B}$ term and which leads to the correct expressions, can be viewed as calculating the changes in the fields as {\it experienced} by the particle, rather than  the changes occurring at a point. It is  plausible and reasonable to at least speculate whether the force acting on a particle might be related to the fields and changes in those fields as experienced by the particle.

The approximations we used reflect a requirement that the gravitation field is weak.  For example, from our expression for the photon deflection, the requirement that the corrections be negligible implies that 
\begin{equation}
\label{MR}
{4\,G\,M\over c^{2}\,R}\gg  \Biggl({4\,G\,M\over c^{2}\,R}\Biggr)^{3}\;\rightarrow \;M\ll {c^{2}\,R\over 4\,G}
\end{equation}
When this condition  is satisfied the corrections to our results (\ref{deflection}) and ({\ref{k2}) are negligible and we are able to reproduce the results of General Relativity. In terms of the length scales the above requirement reads that
\begin{equation}
\label{schw}
R\gg R_{S}
\end{equation}
where $R_{S}=2\,G\,M/c^{2}$ is the Schwarzschild radius for a source with mass $M$. When (\ref{schw}) is not satisfied, our results differ from those of General Relativity; e.g. the expansion of $\arctan \,4\,G\,M/c^{2}\,d$ in (\ref{deflection}) generates the correct functional form for the next correction photon deflection angle, namely  $(4\,G\,M/c^{2}\,d)^{3}$, but with the wrong coefficient $-1/3$. The same is true for the precession of the perihelion.
This higher-order discrepancy is not surprising since our formulation is not proposed as a replacement for, or an improvement of General Relativity. Rather it should be viewed as an attempt to describe the gravitational phenomena using Maxwell-like system of equations. An advantage of the suggested formulation is that  in a leading approximation it reproduces the correct expressions for observables using simple equations (\ref{equations}) - (\ref{!!}), without invoking tensor calculus.

Though the results (\ref{deflection}) and (\ref{k2}) were obtained from General Relativity years ago, it is quite unexpected that they can be obtained using an approach other than one based on Einstein's equations. This alternative ``heuristic'' derivation, and not the expressions (\ref{deflection}) and (\ref{k2}) themselves,  is what we consider as the main result of our paper. From our viewpoint, the possibility of an alternative derivation of expressions for observables such as photon deflection and perihelion advance may provide an impetus for further understanding and insights about the theory of spatial gravitational fields and General Relativity. 

To conclude, from the equations (\ref{equations}) - (\ref{!!})
it is possible to obtain the correct results (in leading approximation in $R_{S}/R$) for the photon deflection angle and the precession of the periastron.

\newpage
\section*{Appendix. Derivation of Equations (\ref{xdzala}) and (\ref{3}) from General Relativity}
Let us sketch how eqs (\ref{xdzala}) and (\ref{3}) can be derived from the equations of General Relativity. These equations also follow from our formulation, and lead to the correct expression for the precession of the perihelion.

We start from the isotropic metric expressed in geometrized units as
\begin{equation}
\label{metric}
g_{00}\,\approx\,-\,\Biggl( 1\,-\,{2M\over R}\,+\,{3M^{2}\over 2R^{2}}\Biggr),\qquad g_{ij}\,\approx\,\delta_{ij}\,\Biggl( 1\,-\,{2M\over R}\Biggr)^{-1}
\end{equation}
Metric (\ref{metric}) is obtained by applying coordinate transformation $R\rightarrow r\,(1\,-\,3M/4r)$ to a Schwarzschild solution $g^{S}_{\mu\nu}(r)$. In the order of approximation used, terms of higher order in $g_{Ij}$ are negligible.

The motion on geodesics is described by
\begin{equation}
\label{geo}
p^{\alpha}p^{i}_{,\alpha}\,+\,\Gamma^{i}_{\alpha\beta}p^{\alpha}p^{\beta}\,=\,0
\end{equation}
where $p^{\alpha}$ is the $\alpha$ component of the momentum, $p^{i}_{,\alpha}\equiv \partial p^{i}/\partial x^{\alpha}$, and $\Gamma^{i}_{\alpha\beta}$ are the Christoffel symbols \cite{landau, dirac}.

For the $x$ - component we have 
\[
m_{0}\,{dp^{x}\over d\tau}\,+\,\Gamma^{x}_{\alpha\beta}p^{\alpha}p^{\beta}\,=\,0
\]
where $\tau$ is affine parameter of the geodesic. The non vanishing Christoffel symbols are 
\[
\Gamma^{x}_{00}\,=\,\Biggl( 1\,-\,{M\over 2R}\Biggr)\,{M\over R^{2}},\qquad \Gamma^{x}_{xx}\,=\,-\Biggl( 1\,-\,{2M\over R}\Biggr)^{-1}\,{M\over R^{2}}
\]
i.e.
\begin{equation}
\label{p0}
m_{0}\,{dp^{x}\over d\tau}\,=\,-\,(p^{0})^{2}\,{M\over R^{2}}\,\Biggl( 1\,-\,{M\over 2R}\Biggr)\,+\,(p^{x})^{2}\,{M\over R^{2}}\,\Biggl( 1\,-\,{2M\over R}\Biggr)^{-1}
\end{equation}
Neglecting higher order terms of  $M/R$ we have $(p^{0})^{2}\,\approx\,m^{2}_{0}\,(1\,+\,2M/R\,+\,{\bf p}^{2}/m^{2}_{0})$ which when substituted into (\ref{p0}) leads to the equation of motion in the $xy$ plane:
\begin{equation}
\fl
\label{p00}
m_{0}\,{dp^{x}\over d\tau}\,=-m^{2}_{0}\Biggl( 1+{3M\over 2R}+{{\bf p}^{2}\over m^{2}_{0}}\Biggr){M\over R^{2}}+{(p^{x})^{2}M\over R^{2}}=-{m^{2}_{0}M\over R^{2}}\,\Biggl (1+{3M\over 2R}\Biggr)\,-\,{M\over R^{2}}\,(p^{y})^{2}
\end{equation}
Using ${\bf p}\,=\,m{\bf v}$ and neglecting higher order terms of $v^{2}/c^{2}$ leads to
\begin{equation}
\label{p01}
{dp^{x}\over d\tau}\,=\,-\,{m_{0}M\over R^{2}}\,\Biggl (1\,+\,{3M\over 2R}\Biggr)\,-\,{m_{0}M\over R^{2}}\,(v^{y})^{2}
\end{equation}
and, using $d\tau\,=\,\sqrt{-g_{00}}\,dt\,\approx \,1\,-\,M/R$,
\begin{equation}
\label{p02}
{dp^{x}\over dt}\,=\,-\,{m_{0}M\over R^{2}}\,\Biggl (1\,+\,{M\over 2R}\Biggr)\,-\,{m_{0}M\over R^{2}}\,(v^{y})^{2}
\end{equation}
Recognizing that $\gamma\approx 1\,+\,M/2R$, $R\equiv u^{-1}$, $m=\gamma m_{0}$, and $v^{y}=u^{-1}\dot{\theta}$ (at $\theta=0$), we see that
equation (\ref{p02}) is the same as our equation (\ref{xdzala}).

Equation ({\ref{3}) is obtained in the same manner, considering the $y$ component of  equation (\ref{geo}).

\section*{References}
\begin{thebibliography}{100}
\bibitem{holz}
Holzmuller G 1870 {\it Z. Math. Phys.} {\bf 15} 6; Tisserand F 1872 {\it Compt. Rend.} {\bf 110} 760
\bibitem{landau}
Landau L D and Lifshits E M 1962 {\it The Classical Theory of Fields} (Oxford: Pergamon Press)
\bibitem{dirac}
Dirac P A M 1976 {\it General Theory of Relativity} (New York: Willey-Interscience)
\bibitem{cataneo}
 Cattaneo C 1958 {\it Nuovo Cimento} {\bf 11} 733
\bibitem{bonnor}
 Bonnor W 1995 {\it Classical and Quantum Gravity} {\bf 12} 499
\bibitem{dunsby}
 Dunsby P K, Basset B A C and Ellis G F R 1997 {\it Classical and Quantum Gravity} {\bf 14} 1215
\bibitem{maartens}
 Maartens R, Ellis G F R and Siklos S T C 1997 {\it Classical and Quantum Gravity} {\bf 14} 1927
\bibitem{mashoon} 
Mashoon B, McClune J and Quevedo H 1997 {\it Physics Letters} {\bf A 231} 47
\bibitem{fj}
 Flanders W D and Japaridze G S 2002 {\it International Journal of Theoretical Physics} {\bf 41} 541
\end{thebibliography}
\end{document}